\documentclass[useAMS,usenatbib]{mn2e}
\usepackage{graphicx}
\usepackage{natbib}
\usepackage{amsmath}
\title[MOST photometry of FU~Ori and Z~CMa]
{Photometric variability in FU~Ori and Z~CMa as observed by 
{\it MOST\/}\thanks{Based on data from: 
\newline 
- the MOST satellite, a Canadian Space Agency mission, jointly operated 
by Dynacon Inc., the University of Toronto Institute of Aerospace 
Studies, and the University of British Columbia, with the assistance
of the University of Vienna, and 
\newline
- the Mount Suhora Observatory, Cracov Pedagogical University.}
}
\author[M. Siwak et al.]
{Michal Siwak$^{1}$\thanks{E-mail: siwak@oa.uj.edu.pl},
Slavek M.\ Rucinski$^2$,
Jaymie M.\ Matthews$^3$,
Rainer Kuschnig$^{3,8}$,
\newauthor
David B.\ Guenther$^4$,
Anthony F.\ J.\ Moffat$^5$,
Jason F.\ Rowe$^6$,
Dimitar Sasselov$^7$,
\newauthor
Werner W.\ Weiss$^8$\\
$^1$Mount Suhora Astronomical Observatory, Cracov Pedagogical University,
ul.\ Podchorazych 2, 30-084 Krakow, Poland\\
$^2$Deparntment of Astronomy and Astrophysics,
University of Toronto, 50 St.\ George St., Toronto,
Ontario, M5S~3H4, Canada\\
$^3$Department of Physics \& Astronomy, University of
British Columbia, 6224 Agricultural Road, Vancouver, B.C., V6T~1Z1, Canada\\
$^4$Institute for Computational Astrophysics,
Department of Astronomy and Physics,
Saint Marys University, \\
Halifax, N.S., B3H~3C3, Canada\\
$^5$D\'{e}partment de Physique, Universit\'{e}
de Montr\'{e}al, C.P.6128, Succursale: Centre-Ville,
Montr\'{e}al, QC, H3C~3J7, Canada\\
$^6$NASA Ames Research Center, Moffett Field, CA 94035, USA\\ 
$^7$Harvard-Smithsonian Center for Astrophysics,
60 Garden Street, Cambridge, MA 02138, USA\\
$^8$Universit\"{a}t Wien, Institut f\"{u}r Astronomie, 
T\"{u}rkenschanzstrasse 17, A-1180 Wien, Austria\\
}

\date{Accepted ;  Received ; in original form }

\pubyear{2013}

\begin{document}
\label{firstpage}
\maketitle

\begin{abstract}
Photometric observations obtained by the {\it MOST\/} satellite were used
to characterize optical small scale variability of the young stars
FU~Ori and Z~CMa. Wavelet analysis for FU~Ori reveals the possible existence
of several 2--9~d quasi-periodic features occurring nearly simultaneously;
they may be interpreted as plasma parcels or other localized disc heterogeneities
revolving at different Keplerian radii in the accretion disc. Their
periods may shorten slowly which may be due to spiralling in of 
individual parcels toward the inner disc radius, estimated at 
$4.8\pm0.2$~R$_\odot$. 
Analysis of additional multicolour data confirms the previously obtained 
relation between 
variations in the $B-V$ colour index and the $V$ magnitude. 
In contrast to the FU~Ori results, the oscillation spectrum of 
Z~CMa does not reveal any periodicities with the wavelet spectrum 
possibly dominated by outburst of the Herbig~Be component.
\end{abstract}


\section{Introduction}
\label{intro}

This paper presents results obtained for two well known young stars,
FU~Ori and Z~CMa, as a continuation of the {\it MOST\/} 
satellite photometric variability studies of Young Stellar Objects.

{\it FU~Ori} is an object of special interest which has been known since 1937, 
when in a timescale of one year its brightness rose from 16 to 9.5~mag 
(in the photographic blue-band system) 
and then started to decay slowly at the rate of 0.015~mag/yr 
\citep{Wachmann,Kenyon2000}. 
This outburst, and two similar events observed in the early 1970's 
in V1057~Cyg and V1515~Cyg, led to the creation of 
a class of eruptive young stars -- ``FUor'' stars \citep{Herbig77} 
-- currently consisting of about 
20 members (\citealt{Semkov10,Semkov11}, \citealt{Miller}). 
The pre-outurst spectra of FUors obtained for two members of the class,
V1057~Cyg and V2493~Cyg (HBC~722), revealed that 
the progenitors were Classical T Tauri- type stars (CTTS).

An enhanced accretion in the disc at a rate of about
$10^{-4} M{_\odot}/yr$, resulting in a major increase of 
the disc surface brightness and dominating over the stellar 
flux, was proposed as the source of the
FUor outbursts \citep{Hartmann85,Hartmann96}. 
Discovery of companions to FU~Ori, Z~CMa and other FUors (\citealt{Wang04, RA04}), 
has initiated discussion whether FUor outbursts may be triggered 
by perturbations in the accretion discs at close periastron passage 
\citep{Bonnell, RA04}.\newline
During an outburst of a FUor, 
emission lines, which are typical for CTTSs, almost completely disappear
while the visual spectrum is dominated by absorption features produced 
in an inner accretion disc radiating as a stellar atmosphere 
of a F-G supergiant star. The outer, colder parts of the FUor 
discs produce a K-M type supergiant 
spectrum observable in the infrared \citep{Kenyon88}. 
In accordance with the location of the brightest 
parts of the Keplerian disc seen in different wavelengths, 
the absorption line broadening diminishes from the visual 
to the near- and mid-infrared spectral regions, as observed 
for FU~Ori by \citet{KH89}, \citet{Hartmann04} and \citet{Zhu09}.

{\it Z~CMa} was recognized as a young star by \citet{Herbig60}. 
The discovery, using infrared speckle interferometry, 
that the star is a very close 0.1 arcsec visual binary \citep{Koresko91,Hass93}, 
resolved the initial difficulties in interpretation of the complex spectral 
properties of the star \citep{Covino}. 
Spectropolarimeric observations revealed that it is the Herbig~Be component 
of the
binary which dominates the infrared continuum and total luminosity 
of the system. The same star is apparently the source 
of the emission lines polarized in the dusty disc envelope and 
observed in visual part of the spectrum \citep{Whitney93}. 
The Herbig~Be star is also responsible for $\Delta V = 1-3$~mag 
outbursts \citep{Ancker04, Grankin09}, caused by variable 
scattering geometry \citep{Szeifert10}, periods of strong mass loss from the 
disc \citep{Benisty10}, or the EXor outburst \citep{Whelan10}. 
These outbursts are superimposed on the slowly decaying light curve 
of the visually brighter FUor star. 
This component contributes a typical -- for FUors -- rotationally 
broadened absorption spectrum produced in the accretion 
disc \citep{Welty92}; it supplied about 80\% of the observed unpolarized 
flux at visual wavelengths \citep{Whitney93} when the system brightness 
was $V=9.7$~mag.

\citet{Kenyon2000} argued that similarly to 
cataclysmic variables, accretion discs of FUors should 
produce flickering variability as a characteristic signature 
observable at visual wavelengths. 
By comparing a limited amount of ground-based 
data to the {\it Monte-Carlo\/} synthetic variability model 
the authors suggested the possibility of $\sim 1$~d 
quasi-periodicities with a $V$-band amplitude of about 0.035~mag, which would originate 
in the inner edge of the FU~Ori disc. 
Motivated by these results and by the lack of similar studies for Z~CMa,
we decided to re-examine this issue 
by means of continuous, high precision, space-based photometric observations. 
Additionally, to obtain information on the {\it wavelength-amplitude} 
dependence of the flickering, 
we observed FU~Ori simultaneously by means of a ground-based 
telescope using intermediate-width Str{\"o}mgren $v$ and $b$ filters. 
We describe details of these observations in Section~\ref{obs}. 
The methods used for light curve analysis and 
the results obtained for our targets are presented 
in Section~\ref{results} and then summarized in Section~\ref{summary}.

\section{Observations and data reductions}
\label{obs}

The optical system of the {\it MOST\/} satellite consists 
of a Rumak-Maksutov f/6, 15~cm reflecting telescope. 
The custom broad-band filter covers the spectral 
range of 380 -- 700~nm with the effective 
wavelength falling close to the Johnson $V$ band.
The pre-launch characteristics of the mission are 
described by \citet{WM2003} and the initial 
post-launch performance by \citet{M2004}.

\begin{figure*}
\centerline{%
\begin{tabular}{c@{\hspace{0pt}}c}
\includegraphics[height=85mm,angle=-90]{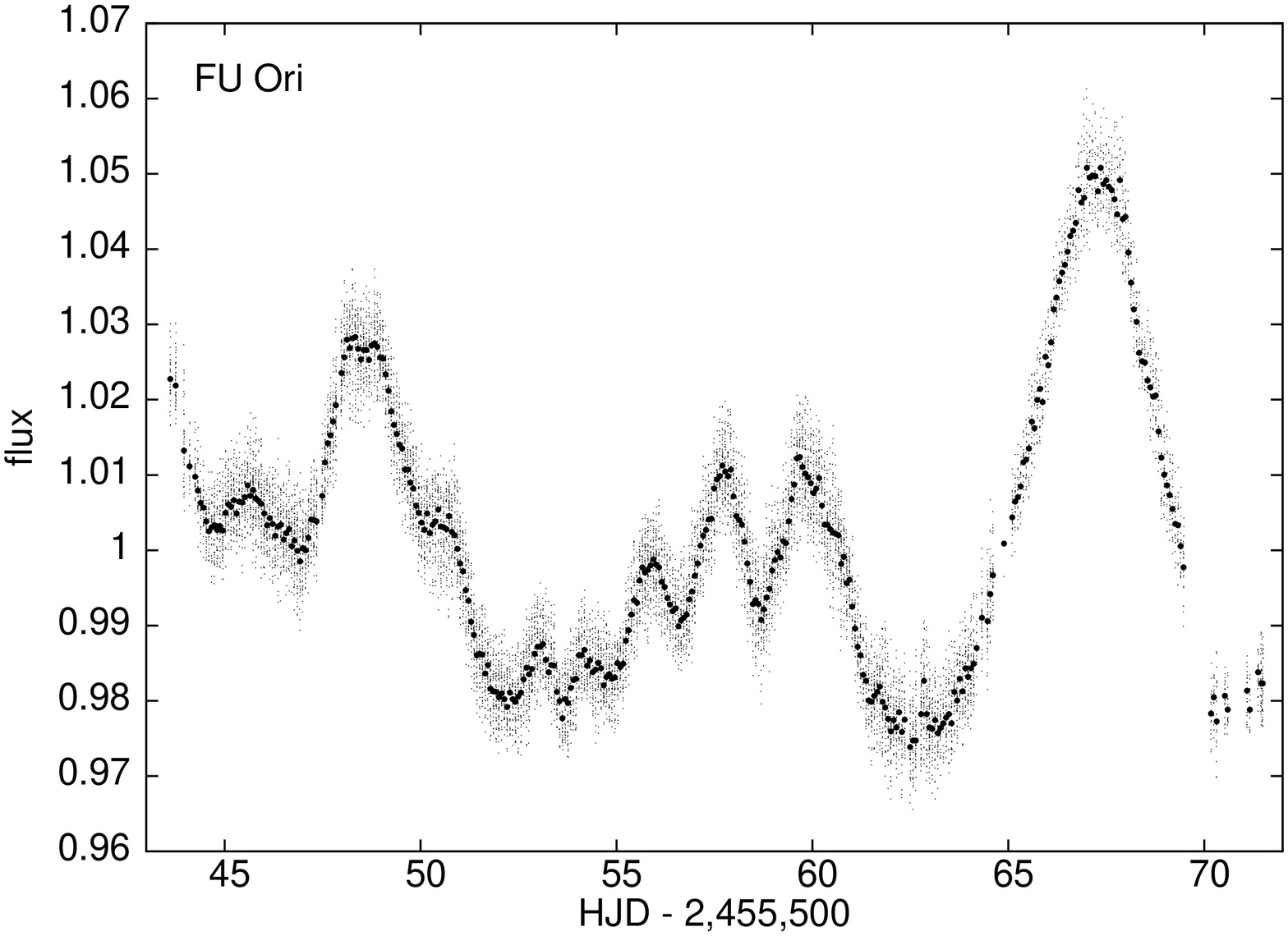}
\includegraphics[height=85mm,angle=-90]{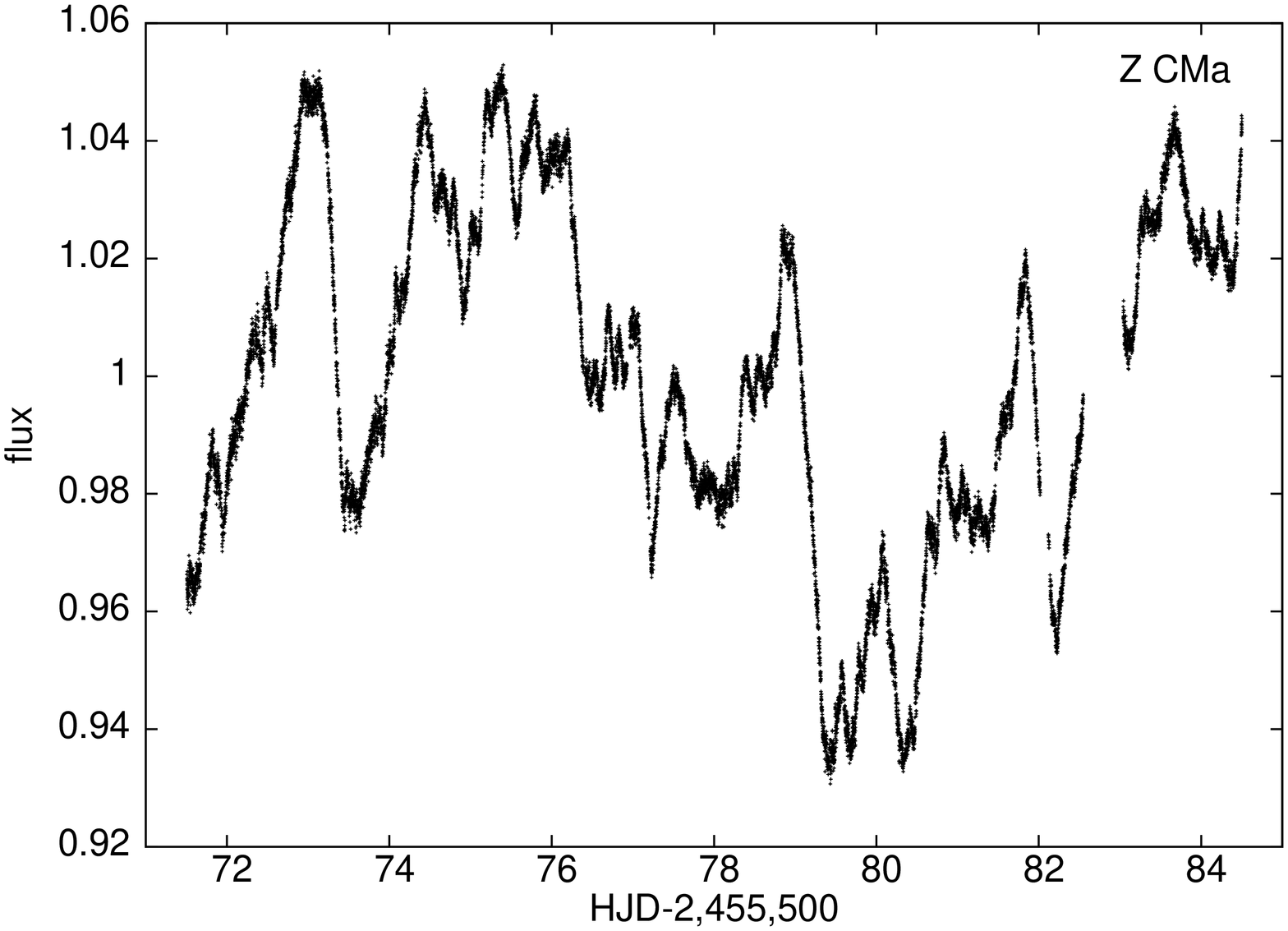} 
\end{tabular}}
\caption{{\it MOST} light curves of FU~Ori (all data points plus 362 {\it mean-orbital} 
averages, left panel) and Z~CMa (all 15404 data points, right panel) 
in normalized flux units.}
\label{Fig.1}
\end{figure*}

The stars investigated in this paper were 
observed in the direct-imaging mode of the satellite. 
FU~Ori was observed nearly continuously for 28 days 
between 13 December, 2010 and 9th January, 2011, 
during 362 satellite orbits. 
The individual exposures were 30~s during the first part 
of the run and 60~s during the last 10 days of the run.
Immediately after the FU~Ori observations, on 10 January 2011 {\it MOST\/} 
started a 13 day long monitoring of Z~CMa, which was observed 
with 60~s exposures. 
Some occasional interruptions in data acquisition, 
visible in the light curves (Figure~\ref{Fig.1}) did not impact 
the scientific results.

The {\it dark} and {\it flat} calibration 
frames for the {\it MOST\/} data were obtained by averaging 
a dozen {\it empty-field} images specifically 
taken during each observing run, or -- for the case of the 
30~sec long exposures of FU~Ori -- 
from frames with the target localized far beyond its optimal position 
due to occasional satellite guiding errors.
Aperture photometry of the stars was obtained 
from the {\it dark} and {\it flat} corrected 
images by means of the DAOPHOT~II package \citep{stet}.
As in our previous investigations, a weak correlation between the star flux 
and the sky background level within each {\it MOST\/} orbit
was noted and removed; 
it was most probably caused by a small photometric nonlinearity in 
the electronic system (see \citealt{siwak}). 
As a result, we obtained very good quality light curves,
particularly for Z~CMa (see Fig.~\ref{Fig.1}).
In the analysis we used the satellite-orbit (101 min) averages for FU~Ori
whose median error was 0.0036 of the mean normalized flux. 
We note that FU~Ori is surrounded by a bright nebula which
slightly decreased the photometric accuracy of the data.
For the brighter Z~CMa, the data were binned into smaller, 
14.5~min bins consisting 
of typically 10 -- 15 observations with the median error 0.0011 per bin.
The finer binning was done in order to preserve 
information on the short time-scale variability of the star. 

During four nights of Jan 4, 5, 8, 9, 2011, 
we simultaneously observed FU~Ori in $v$ and $b$ Str{\"o}mgren 
filters using the 60~cm telescope at the Mount Suhora Observatory, Cracow 
Pedagogical University. We were unable to obtain similar data for Z~CMa
due to poor weather conditions.
The data were reduced in a standard way in the ESO-MIDAS software environment. 
Aperture photometry was obtained with the DAOPHOT~II package. 
The photometry of FU~Ori was done differentially utilizing 
a {\it mean comparison star} made of the three nearby, somewhat fainter 
stars having colour indices similar to FU~Ori (Tab.~\ref{Tab.1}).
Thanks to the proximity of the stars and similarity of the 
colours, corrections for differential 
and colour extinction were unnecessary.

\begin{table}
\caption{Basic informations on FU~Ori and the
comparision stars used for differential photometry.
\newline 
*The average $V$ magnitude of FU~Ori
during {\it MOST\/} observations as obtained from observations made by 
Konstatin Grankin (priv.comm.), Bruno Alain and Timar Andras (AAVSO).}
\begin{tabular}{c c c }
\hline
star             & V                 & B-V        \\ \hline
FU~Ori           & 9.74$^*$          & 1.279(47)  \\
GSC~00714-0203   & 10.555(88)        & 1.069(85)  \\ 
GSC~00715-0188   & 10.682(93)        & 1.646(241) \\
GSC~00715-0123   & 11.337(119)       & 0.960(194) \\ \hline
\end{tabular}
\label{Tab.1}
\end{table}

\begin{figure*}
\centerline{%
\begin{tabular}{c@{\hspace{0pt}}c}
\includegraphics[height=85mm,angle=-90]{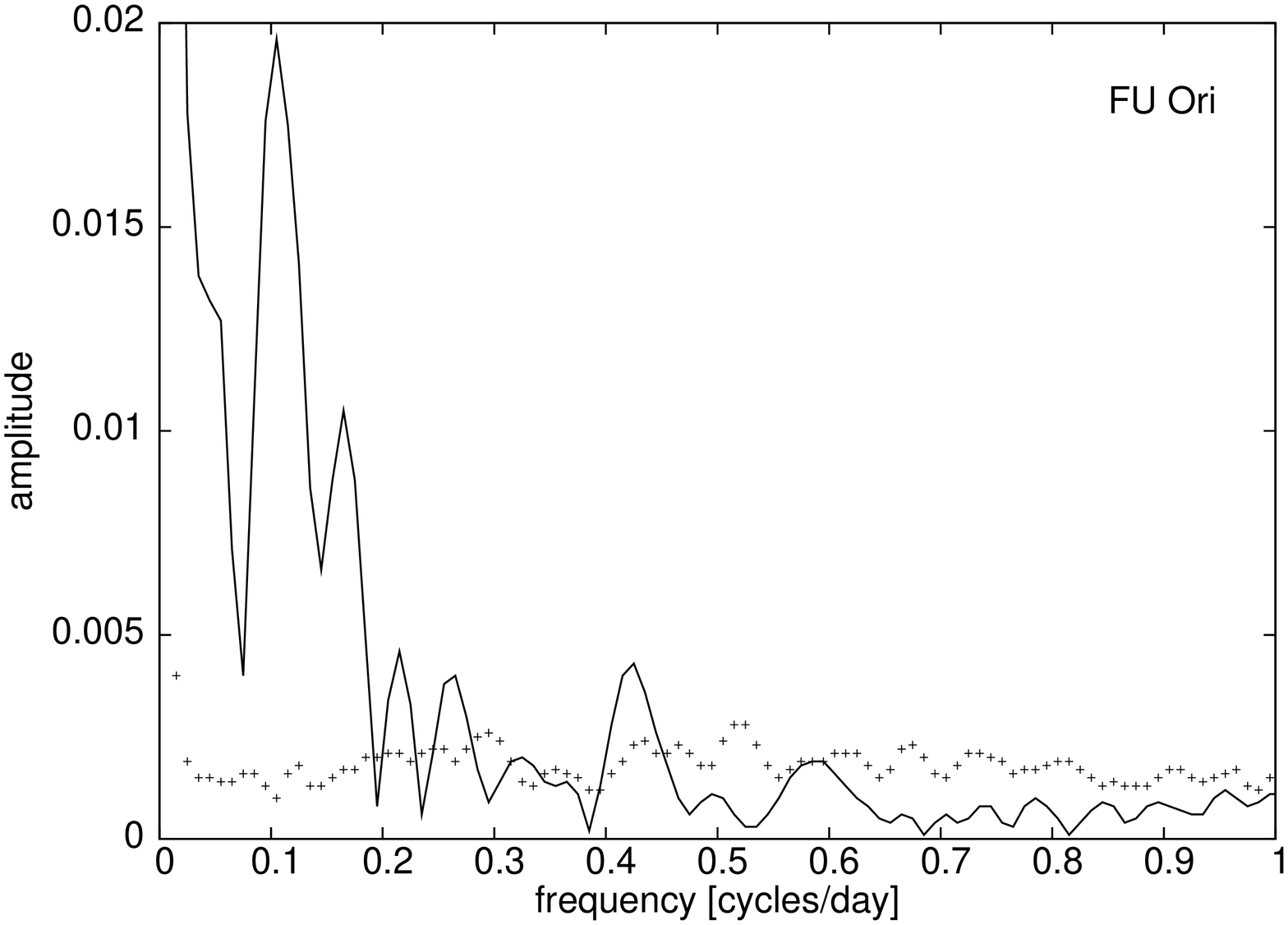}
\includegraphics[height=85mm,angle=-90]{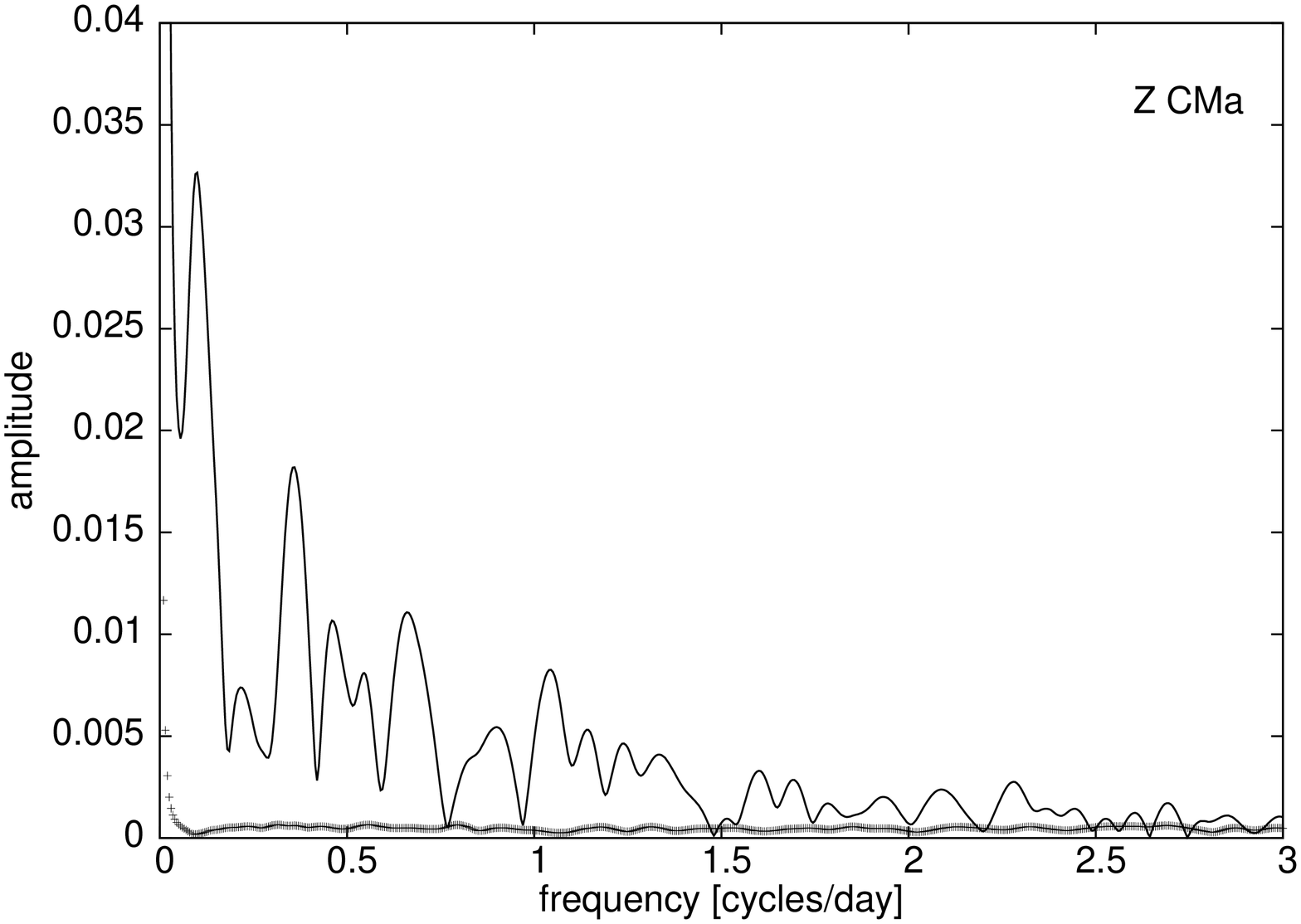} 
\end{tabular}}
\caption{Fourier analysis of the variability data 
computed from the {\it mean-orbital} data points of FU~Ori 
(left panel) and from all data points of Z~CMa (right panel). 
The amplitude errors estimated through bootstrap 
repeated sampling are represented by small points.}
\label{Fig.2}
\end{figure*}

\begin{figure*}
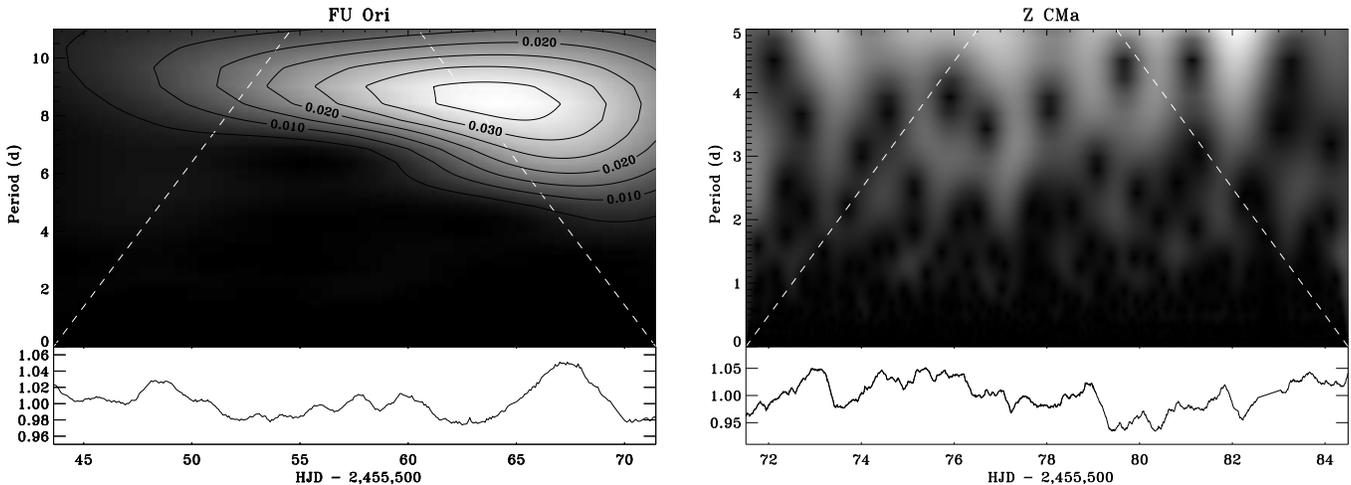

\centerline{%
\begin{tabular}{c@{\hspace{0pt}}c}
\includegraphics[height=65mm,angle=0]{figures/fuori11pow.eps}
\includegraphics[height=65mm,angle=0]{figures/zcmaall.eps} 
\end{tabular}}
\caption{Morlet-6 wavelet spectra of FU~Ori (left panel) and Z~CMa (right panel) calculated 
for the whole accessible period range up to 11 and 5 days, respectively. 
The light curve in normalized flux units is shown at the bottoms of the panels. 
Ranges not affected by edge effects in the 
wavelt transformation are located between the two white broken lines.}
\label{Fig.3}
\end{figure*}

\section{Results of the light-curve analysis}
\label{results}

We performed analysis of the {\it MOST\/} data in a similar way 
to \citet{ruc08,ruc10} and \citet{siwak2}, i.e. 
the Fourier analysis was done by simple least-squares fits 
of expressions of the form $l(f ) = c_0 (f) + c_1 (f)cos[2\pi (t - t_0) f ] + c_2 (f ) sin [2\pi (t - t_0) f ]$ 
for an appropriate range of frequencies $f$, with the step of $\Delta f = 0.01$. 
The amplitude a(f) for each frequency was evaluated as the modulus
of the periodic component, $a(f)=\sqrt{c_1^2 (f) + c_2^2(f)}$.
The bootstrap sampling technique permitted evaluation of mean standard
errors of the amplitudes from the spread of the coefficients $a_i$. 
This technique, for a uniform temporal sampling -- as in our case -- may
give too pessimistic estimates of errors, but consistently we prefer this conservative 
approach. 

For the Fourier analysis of the FU~Ori variability,
we used 362 {\it mean-orbital} data points, 
while in the case of Z~CMa, which is much brighter,
all 15404 measurements, obtained every 60~sec were used. 
For the wavelet analysis of Z~CMa, the data have been averaged and partly 
interpolated into 1291 points on an equal-spacing grid 
at 0.010067~d (14.5~min) steps.
Similar interpolation for FU~Ori, which suffered more
interruptions, led to mapping into a grid of 395 points. 
The Morlet-6 wavelet provided the best match between the time-integrated 
power spectrum and original frequency spectrum of both FU~Ori and Z~CMa. 
This was noticed by \citet{ruc08} and encountered later 
by \citet{ruc10} and \citet{siwak2}. 
The Morlet transforms of other orders result in systematic differences in the period scale.

\subsection{FU Ori}
\label{fuori}

\subsubsection{Analysis of the Fourier and the wavelet spectrum}

The upper envelope of peak amplitudes in the Fourier transform 
of the FU~Ori data (Fig.~\ref{Fig.2}, left panel) appears 
to scale as flicker noise, $\propto 1/\sqrt{f}$, where $f$ is the frequency.
The first most prominent maximum at $f = 0.107$~c/d 
is visible as the largest and most intense 
feature in the wavelet spectrum (Fig.~\ref{Fig.3}, left panel) and directly 
in the light curve. 
Its period shortens gently from $\sim$9 to $\sim$8 days, 
somewhat similarly to the case of TW~Hya \citep{ruc08,siwak2}. 
However, we note that this periodicity is based on only three
cycles so that its reality may be questioned. To test the validity
of our results, we conducted tests on recovery of drifting, 
quasi-periodic features which additionally change their amplitudes.
In each case we obtained very satisfactory reproduction of synthesised 
light curves, including the rates of period and amplitude changes.\newline
Following our interpretation of TW~Hya, 
we assume that the oscillation feature is produced by plasma 
condensations and/or other accretion disc heterogeneities, 
in turn produced by interactions of stellar magnetic field with 
the inner disc plasma, although we have no clear picture what
causes these instabilities. 
If this is a correct view, then the period variations 
reflect a change in localization within the disc at radii (estimated
from Keplerian periods) from about 12.2 to 11.3~R$_{\odot}$, 
assuming a central star mass of 0.3 M$_{\odot}$ \citep{Zhu07}. 
The low inclination of the accretion disc
of 55~deg \citep{Malbet2005} should indeed permit to observe such 
inner disc structural changes. 

Other peaks in the Fourier spectrum
(Figure~\ref{Fig.2}) localized at f = 0.166, 0.215 and 0.263 c/d, (P = 6.0, 4.7 
and 3.8 d, respectively) are represented by the much fainter and apparently 
constant features visible in different parts of the wavelet spectrum. 
They cannot be well characterized and may actually represent processing 
artifacts for the relatively short dataset or a pure random flickering.  
We note that these three periodicities are not directly visible in 
the FU~Ori light curve.

\begin{figure}
\includegraphics[width=85mm,angle=0]{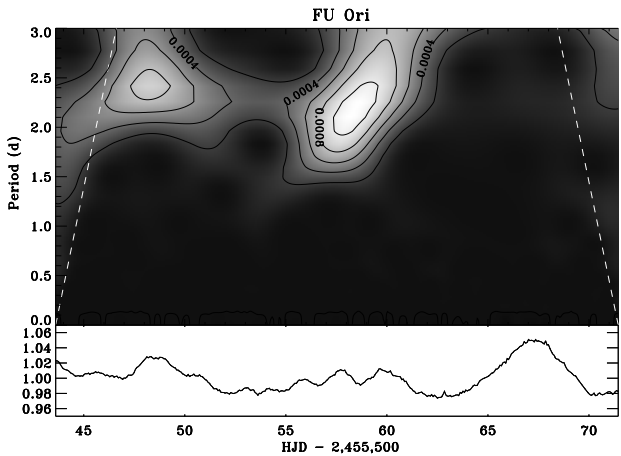}
\caption{The upper panel shows the wavelet spectrum of FU~Ori (as in Fig.~\ref{Fig.3}) but 
limited to 3 days to improve visibility of shorter oscillations. The light curve in normalized 
flux units is shown at the bottom of the panel. Ranges not affected by edge effects in the 
wavelt transformation are located between the two white broken lines.}
\label{Fig.4}
\end{figure}

In addition to the features described above, 
the light curve contains a well defined directly visible short-periodic signal 
at $f=0.424$~c/d, 
which disappears after $HJD\approx2,455,560$. 
To improve its visibility we prepared 
a second wavelet image (Fig~\ref{Fig.4}), which reveals 
one apparently drifting feature with two well-defined peaks: 
the first one is localized at $P=2.4$~d and the second at $P=2.2$~d. 
Within the model of orbiting plasma in the inner disc
this would correspond to the Keplerian radii within 5.1 to 4.8 R$_\odot$. 
We do not see any obvious short-period signals at periods 
shorter than $P=2.2$~d which may indicate that the 
inner accretion disc is truncated at about 4.8 R$_\odot$.
This value is in agreement with the inner disc 
radius of 5.5$^{+2.9}_{-1.8}$~R$_{\odot}$ obtained from 
interferometric observations by \citet{Malbet2005} 
but is 3 times smaller from that obtained by \citet{Eisner} -- $15\pm4$~R$_{\odot}$.

\subsubsection{Analysis of multicolour observations}

\begin{figure}
\includegraphics[width=55mm,angle=-90]{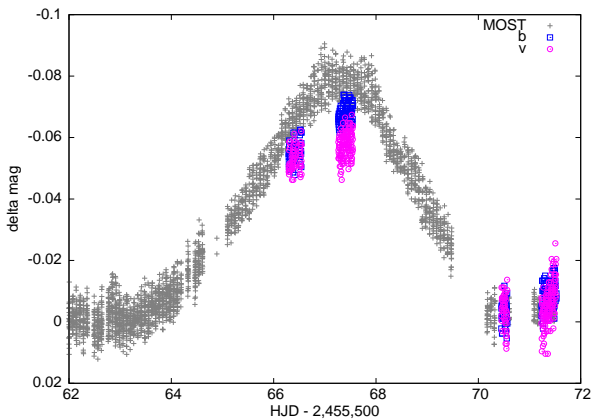}
\caption{A fragment of FU~Ori light curve covered by the
ground-based data. The light curves from MOST (small crosses) and in
$b$ and $v$ filters (squares and circles) were shifted to the same level 
at the light minimum for an easier comparison.
}
\label{Fig.5}
\end{figure}

Although the weather conditions during the
ground-based observations were poor, we were very fortunate 
to have the four clear nights exactly when the star 
showed both the maximum and the minimum of its 8~d long 
oscillation (Fig~\ref{Fig.5}). 
This enabled us to address the {\it wavelength-amplitude} 
dependence in the $b$ (4670\AA) and $v$ (4110\AA) filters 
of the Str{\"o}mgren photometry.

As discussed by \citet{Kenyon2000}, the observed variability
amplitudes decrease at shorter wavelengths. 
In order to compare our results with the relation
of the changes $\Delta(B-V)$  and $\Delta V$ found by \citet{Kenyon2000}, 
we used the average magnitudes in our two filters and found the mean
colour index $\Delta(0.5\times[v+b])-\Delta V = 0.016 \pm 0.002$~mag.
It is larger by 0.06~mag than that returned 
by Equation~6 of the \citet{Kenyon2000} paper, but agrees with their finding 
that the $B-V$ colour index becomes redder as the star becomes brighter. 
This relation allowed the authors constrain the effective spectrum 
producing the flickering to F7--G3 spectral type. 

The 0.06~mag difference in the colour index 
can be explained by the difference in the location of the
dominant source of the observed stellar flux. The respective
annuli in the disc, at different Keplerian distances,
should produce different {\it wavelength-amplitude} relations.
The 9--8~d oscillation feature is assumed to be produced 
at the distance of $12-11~R_{\odot}$, where according to FU~Ori disc 
models \citep{Zhu07}, $T_{eff}\approx1500$~K, so that
the maximum energy is located at $\lambda\approx10\mu m$. 
Obviously, the amplitudes of flux variations 
arising at such a large distance are considerably 
reduced by the dominant flux produced in the innermost disc.
As the feature approaches the central star on a
spiral orbit, it moves to warmer disc annuli. 
As a consequence, the wide-band {\it MOST\/} filter
receives initially a small fraction of the energy emitted  
at 12~R$_{\odot}$, but then this fraction increases
as the spiralling-in process continues. 

Changes in the amplitude of the 9--8 day periodic feature
may be questioned in view of the duration of the 
whole {\it MOST\/} observations lasting 4 weeks. 
However, the general trend of the progressively growing amplitude
as the period shortens seem to be visible both, directly in the light curve 
(Fig~\ref{Fig.1}, left panel) 
and in the wavelet spectrum (Fig~\ref{Fig.3}, left panel).
The amplitude  reached about 0.03~mag in 
the first and the second cycle (the overlapping non-sinusoidal oscillations 
do not allow for a more accurate estimate), and 0.075~mag in the third. 
Once again we stress, that this finding bases on three 
oscillations only. 
However, our data do reveal a similar effect, of the amplitude growing 
up from 0.005~mag to 0.02~mag, also for the $P=2.4-2.2$~d feature; such a drift 
would correspond to a motion  
from 5.1 to 4.8~R$_\odot$, where the disc annuli 
have temperatures $T_{\rm eff}\approx3400$~K \citep{Zhu07}.
The difference in the amplitudes with \citet{Kenyon2000} may have resulted 
from averaging of independent wave trains during their observations or may reflect 
a real, physical difference in sizes of the gas elements.

\subsection{Z CMa}
\label{zcma}

As mentioned in Section~\ref{intro}, about 80\% of the
unpolarized visual flux of Z~CMa is produced by the FUor component.
The direct flux from the Herbig~Be star is obscured by a
dusty envelope and visible as scattered continuum light, which is 
polarized at about 6\% and as emission lines \citep{Whitney93}.
The {\it MOST\/} satellite does not have any polarization capabilities 
and measures the total flux. 
The observations in January 2011 were obtained during the light maximum 
of a new outburst when, according to the AAVSO database, the average total 
system brightness was $V=8.7$~mag. 
This may explain the chaotic shape of the wavelet transform (Fig.~\ref{Fig.3}, right panel) 
containing many incoherent variations down to time scales of about 0.3 day.
At least a part of these oscillations may have their origin in the night-to-night 
and hour-to-hour variations of H$_\alpha$, H$_\beta$, Na~I D P~Cygni absorptions 
and in the H$_\beta$ emission peak \citep{Chochol}. 
Also, from speckle observations obtained in the infrared $H\&K$ 
filters \citet{Hass93} reported that both components of the
close visual double were varying independently by $\sim0.5$~mag.

The time variability spectrum of Z~CMa is somewhat 
similar to that of the Herbig~Ae star HD~37806 \citep{ruc10}. 
The chaotic variability at all time scales and lack of
any regularity in the Fourier or wavelet transforms
strongly suggest that the time variability spectrum is dominated
by the Be component which contributes at least 20\% to the total
flux. Lack of any firm results is compounded by
the relatively short duration of the observation run which lasted 
only 13 days.
Unfortunately, the {\it MOST\/} observations of Z~CMa done one year earlier, in 
2010, obtained in a time of relatively moderate light
variations were of a very poor quality due to technical problems.

\section{Summary}
\label{summary}

Continuous {\it MOST\/} satellite observations of FU~Ori and Z~CMa 
confirmed presence of intensive, rapid (time scales of single days) 
variations of their light which may have general characteristics
of the flicker noise with amplitudes scaling as: $a \propto 1/\sqrt{f}$,
similar to the Classical T~Tauri star TW~Hya \citep{ruc08,siwak2}.
For FU~Ori, the Fourier and wavelet transform spectra show 
quasi-periodic features, one well defined and possibly changing
its period within 9 to 8~day, and one much fainter, at about
2.4~days drifting down to 2.2~days. 
But we note an absence of any 1-day periodicity in our data,
contrary to the \citet{Kenyon2000} results which were obtained from 
a limited amount of unequally-spaced, poor-quality data. 
We tentatively interpret these quasi-periodic variations as 
produced by hot plasma condensations and/or disc heterogeneities 
which develop in the magneto-rotationaly unstable inner parts at the 
Keplerian distances of about 12 and 5~R$_\odot$. The period
variability may be interpreted as a spiralling-in or inward drifts 
in the innermost disc. 
The shortest observed period of $2.2\pm0.1$~d may
define the inner edge of the accretion disc at $4.8\pm0.2$~R$_{\odot}$,
which agrees with the estimate of  \citet{Malbet2005}
from the interferometric observations. 
Within the {\it disc-locking} mechanism picture of the accretion processes,
the rotational period of the star would be then close to 2 days. 
Although this was the interpretation of such oscillatory features for TW~Hya 
\citep{ruc08,siwak2},
Herbst (priv. comm.) argued that for the case of TW~Hya, 
hot condensation in the disc could not easily 
produce up to 20\% flux variations observed in the visual 
spectrum which is dominated by the central star. This is indeed
a valid point, but we note that for the
case of FU~Ori the accretion disc 
luminosity overwhelms that of the central star by about 100$\times$. 

Two colour Str{\"o}mgren $v$ and $b$ ground-based observations 
of FU~Ori confirmed the $\Delta(B-V)$ versus $\Delta V$ relation obtained 
by \citet{Kenyon2000}. 
However, we obtained slightly redder colour index of the dominant variations, 
what may be due to a more outward hotspot location (causing the 8 d periodicity) 
during our observations. 
This would agree with the interpretation of the {\it wavelength-amplitude} relationship
through different locations of the dominant variable flux,
with longer periods produced by more external and cooler 
parts of the accretion disc.
In generally this effect may manifest itself within the wide-band {\it MOST\/} 
light curve and in the wavelet transform of the star variability
as the amplitudes of the two drifting, oscillatory 
features markedly increased as their periods decreased.
Future simultaneous {\it MOST\/} and ground-based, multi-band observations conducted 
over a few months may help in investigation of the {\it wavelength-amplitude} relation for 
a range of possible periods. 
Such observations should be supported by spectral synthesis disc models, to localize 
the dominant light sources at various inner disc concentric annuli. 

Z~CMa, the complex binary consisting of a FUor and Herbig~Be stars,  
was observed by {\it MOST\/} during 13 days close in time to
the outburst maximum of 
the Herbig~Be component. This led to the suppression of the
FUor variability component and dominance of the 
very chaotic spectrum of oscillations of the Be component. 
The variability spectrum is somewhat similar
to that of the Herbig~Ae star HD~37806 \citep{ruc10}. 
Future observations during quiescence level could help to disentangle 
the two light contributions. 

\section*{Acknowledgments}

MS thanks the Mount Suhora Observatory Staff for the very generous time 
allocation and the hospitality, as well as the Canadian Space Agency 
Post-Doctoral position grant to Prof. Slavek M. Rucinski within the framework 
of the Space Science Enhancement Program.
\newline
The Natural Sciences and Engineering Research Council of
Canada supports the research of DBG, JMM, AFJM, and SMR.
Additional support for AFJM comes from FQRNT (Qu\'ebec).
RK is supported by the Canadian Space Agency and WWW is supported 
by the Austrian Science Funds (P22691-N16). 
\newline
We acknowledge with thanks the two anonymous referees for the very valuable comments 
concerning important issues of this paper, as well as the variable star observations
from the AAVSO International Database contributed by observers worldwide
and used in this research.
This research has made use of the SIMBAD database, operated at CDS, 
Strasbourg, France and NASA's Astrophysics Data System (ADS) Bibliographic 
Services.

\end{document}